# Magneto-optical determination of the carrier lifetime in coherent Ge$_{1-x}$Sn$_x$/Ge heterostructures

Elisa Vitiello[1], Simone Rossi[1], Christopher A. Broderick[2,3], Giorgio Gravina[1], Andrea Balocchi[4], Xavier Marie[4], Eoin P. O'Reilly[2,3], Maksym Myronov[5] and Fabio Pezzoli[1]

1. LNESS and Dipartimento di Scienza dei Materiali, Università degli Studi di Milano-Bicocca, via R. Cozzi 55, I-20125 Milano, Italy
2. Tyndall National Institute, University College Cork, Lee Maltings, Dyke Parade, Cork T12 R5CP, Ireland
3. Department of Physics, University College Cork, Cork T12 YN60, Ireland
4. Université de Toulouse, INSA-CNRS-UPS, LPCNO, 135 Avenue de Rangueil, 31077 Toulouse, France
5. Department of Physics, The University of Warwick, Coventry CV4 7AL, United Kingdom

**We present a magneto-optical study of the carrier dynamics in compressively strained Ge$_{1-x}$Sn$_x$ films having Sn compositions up to 10% epitaxially grown on blanket Ge on Si (001) virtual substrates. We leverage the Hanle effect under steady-state excitation to study the spin-dependent optical transitions in presence of an external magnetic field. This allowed us to obtain direct access to the dynamics of the optically-induced carrier population. Our approach singled out that at cryogenic temperatures the effective lifetime of the photogenerated carriers in coherent Ge$_{1-x}$Sn$_x$ occurs in the sub-ns time scale. Supported by a model estimate of the radiative lifetime, our measurements indicate that carrier recombination is dominated by non-radiative processes. Our results thus provide central information to advance the fundamental understanding of carrier kinetics in this novel direct-gap group-IV material system. Such knowledge can be a stepping stone in the quest for the implementation of Ge$_{1-x}$Sn$_x$-based functional devices.**

Ge$_{1-x}$Sn$_x$ binary alloys hold the promise of introducing novel and advanced functionalities onto the well-established and wide-spread Si technology[1–5]. The recent introduction of out-of-equilibrium growth techniques has been a key-enabling factor to mitigate the longstanding problems associated to the low solubility of Sn in the Ge and Si lattices and to tame defect injection induced by the sizeable lattice mismatch between these elements[6–8]. As a result, epitaxial Sn-containing alloys are now offering direct access to a wealth of intriguing phenomena, hence spurring a surge of interest. Ge$_{1-x}$Sn$_x$/Ge heteroepitaxial structures have recently shown tremendous improvements in lasing performances[9–12] and competitive photodetection efficiencies[13–15] in the technologically important mid-infrared wavelength range. These efforts are bringing Ge$_{1-x}$Sn$_x$ heterostructures a step closer to a practical implementation as off-the-shelf photonic components. The intriguing application of strain and bandgap engineering, a well-established technology in group IV heterostructures[16], could additionally yield the use of Ge$_{1-x}$Sn$_x$ as mobility booster in next-generation microelectronic devices[17,18]. Recently, the potential of Ge$_{1-x}$Sn$_x$ heterostructures in novel fields such as spintronics and quantum technologies has been proposed[19].

Despite these extraordinary advancements, key properties of $Ge_{1-x}Sn_x$ binary alloys relevant to device applications have largely remained unexplored. Notably, there is a critical lack of understanding of the mechanisms dominating carrier dynamics and non-radiative recombination, notwithstanding their crucial implications for transport and optical phenomena. Yet, such knowledge is pivotal also to inform facilitate future implementations of proposed photonic and electronic devices. Time-resolved (TR) methods applied to a vast range of experimental techniques, e.g. absorption and photoluminescence (PL), are a primer to study the kinetics of the carriers. Although successful in addressing the spin physics of $Ge_{0.95}Sn_{0.05}$[19], TR spectroscopy has been seldomly applied to $Ge_{1-x}Sn_x$ alloys, owning to the narrow amplitude of the energy gap. In addition to nontrivial structural problems associated to the spontaneous segregation of Sn, the bandgap of $Ge_{1-x}Sn_x$ typically lies in the shortwave or mid-infrared. This has jeopardized a facile experimental achievement of adequate time and spectral resolutions, leading to a lack of insight regarding carrier dynamics in this emerging material system.

In this work, we introduce a radically different approach to provide such insight. We leverage steady-state optical spin orientation combined with a magneto-optical analysis of the radiative emission of ultrathin $Ge_{1-x}Sn_x$ epitaxial layers. Specifically, we apply a technique so far overlooked in the context of group IV materials, namely the optical Hanle effect in which the PL polarization is suppressed by an external magnetic field[20]. According to previous results for III-V compounds, the resulting polarization decay can be ascribed to the photogenerated carrier dynamics[20]. Here, we exploit the Hanle effect to reveal in $Ge_{1-x}Sn_x$ the presence of an effective carrier lifetime in the sub-nanosecond regime without relying on demanding TR techniques. We demonstrate the effectiveness of the present approach exploring a wide Sn composition range spanning from 2 to 10%. Finally, we utilize this method to gather access to non-radiative recombination induced by plastic strain relaxation, thus obtaining central information, namely the interface recombination velocity at the defective $Ge_{1-x}Sn_x/Ge$ junction.

$Ge_{1-x}Sn_x/Ge/Si(001)$ heterostructures were grown using an ASM Epsilon 2000 industrial-type reduced-pressure chemical vapor deposition (RP-CVD) system. The epilayers of $Ge_{1-x}Sn_x$ were deposited on a 100-mm diameter Si(001) wafer via a relaxed Ge buffer with thickness $h \sim 600\ nm$ [see the schematic in the inset of Figure 1]. The structural properties of the samples are listed in Table 1. All the $Ge_{1-x}Sn_x$ films are under compressive strain and below the epilayer's critical thickness for the corresponding Sn content. A sample without the $Ge_{1-x}Sn_x$ film was also measured as a reference.

Continuous-wave (CW) PL experiments were performed at 10 K by exciting the samples with a Nd-YVO$_4$ laser at 1.165 eV. The spot diameter was about 50 µm and the resulting optical power density several kW/cm$^2$. The polarized PL was dispersed by a monochromator and coupled to an InGaAs photodiode with a cutoff at ~0.5 eV.

Table 1 Structural properties of $Ge_{1-x}Sn_x$ layers grown coherently on the almost relaxed Ge buffer, whose residual biaxial tensile strain stems from the difference in the thermal expansion coefficient with respect to the Si substrate. The thickness h of the layers have been experimentally determined from X-ray diffraction (XRD) data.

| Sn Content (%) | h (nm) |
|---|---|
| 2 | 150 |
| 3 | 45 |
| 4 | 35 |
| 6 | 70 |
| 8 | 40 |
| 10 | 40 |

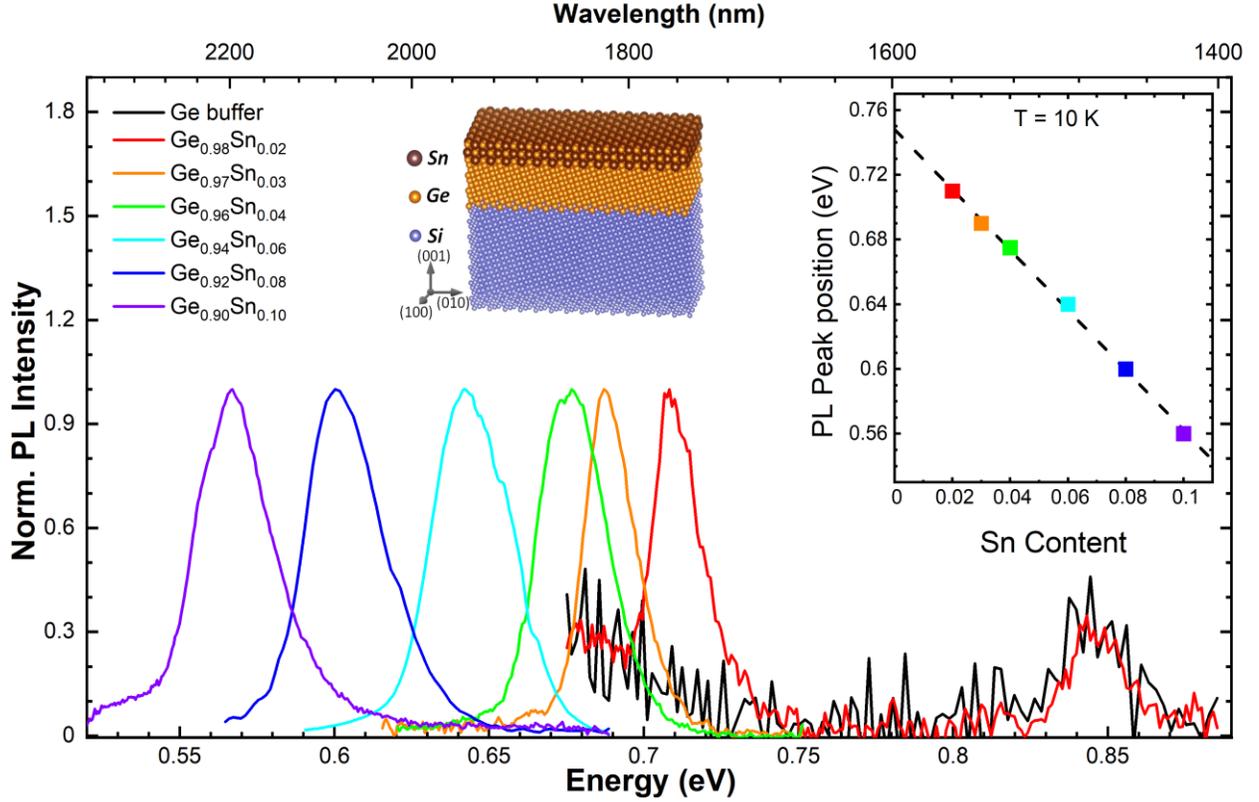

**Figure 1** 10 K PL spectra of $Ge_{1-x}Sn_x$ with x ranging from 0 (Ge buffer, black curve), to 0.1 (violet curve). A sketch of $Ge_{1-x}Sn_x$ /Ge/Si heterostructure is presented: coherent $Ge_{1-x}Sn_x$ layers with various tin content were grown on 600 nm-thick Ge buffer layer deposited on (001) Si substrate. The inset shows the PL peak position as a function of the Sn content. Dashed line is the linear fit.

Figure 1 shows low temperature PL spectra of all the investigated heterostructures, at temperature of 10 K. The tensile thermally strained Ge buffer layer shows a high-energy PL feature at about 0.85 eV (black line in Figure 1), corresponding to direct-gap radiative recombination in Ge[21–23]. The addition of an ultrathin coherent $Ge_{0.98}Sn_{0.02}$ cap layer to the Ge/Si epitaxial stack yields the emergence of a new prominent PL feature at about 0.71 eV (see the red spectrum in Figure 1). Such peak can be ascribed to band-to-band recombination events occurring in the topmost $Ge_{0.98}Sn_{0.02}$ film. Increasing the Sn molar fraction up to x = 10% redshifts the PL peak energy to 0.56 eV, in line with the Sn-induced band gap reduction. The inset of Figure 1 outlines this shrinkage of the bandgap, which is determined by a combination of increasing Sn content in the $Ge_{1-x}Sn_x$ alloy and subsequent increase of compressive strain in the epilayer, as reported previously[24,25]. The linear regression of the data (dashed line in Fig. 1) gives an intercept of about 0.75 eV at x = 0, i.e. elemental Ge. This value is consistent with the energy expected at low temperature for the optical transition through the fundamental indirect gap of bulk Ge[26], thereby suggesting the prominence of the indirect band gap recombination in all the observed spectra of our coherent $Ge_{1-x}Sn_x$ films. The emergence of the indirect PL in $Ge_{1-x}Sn_x$ is also in line with the Γ-L hybridization of the conduction band edge pointed out by recent calculations[27,28].

Each PL peak was then studied through optical spin orientation. The circular polarization degree ($\rho_{circ}$) of the PL emission was measured after circularly polarized laser excitation. Details about this technique can be found in previous works[21,29,30]. On average, $\rho_{circ} \sim 12\%$ in our $Ge_{1-x}Sn_x$ epitaxial films, demonstrating robust optical spin orientation. Since the spin relaxation time of holes is expected to be significantly shorter than that associated with electrons, the latter primarily determines the experimentally accessible $\rho_{circ}$ value[19,31,32].

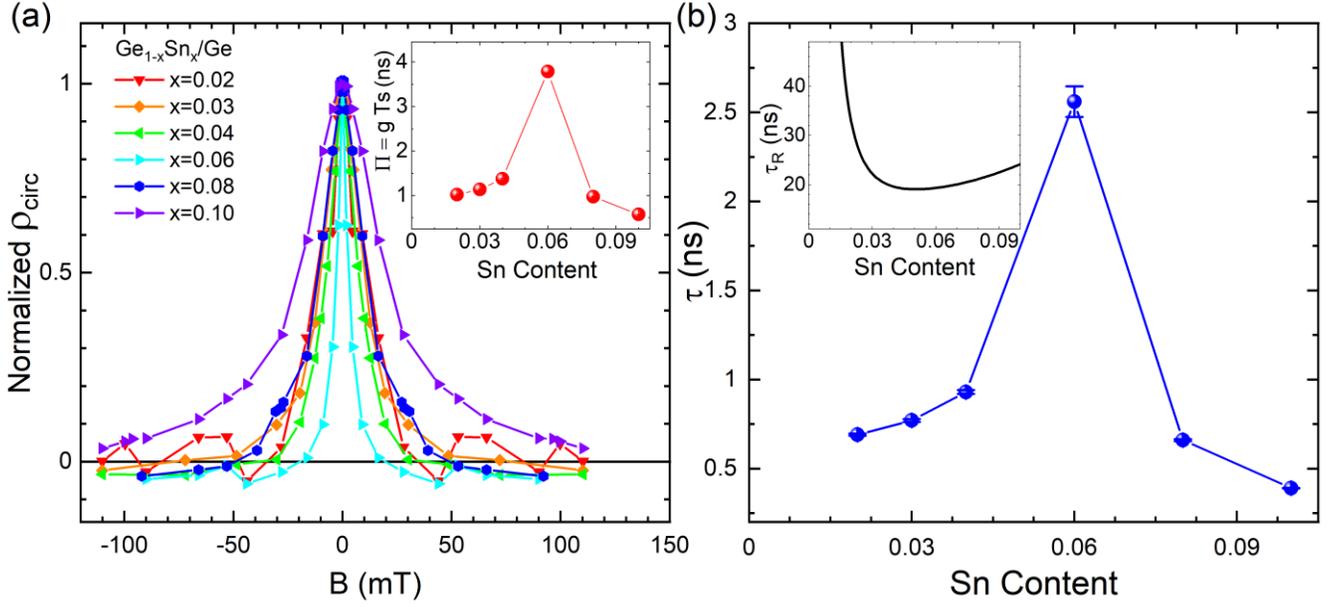

**Figure 2** (a) Low temperature (10 K) circular polarization degree ρ$_{circ}$ as a function of transverse magnetic flux density (Hanle curve), normalized with respect the maximum $\boldsymbol{\rho_{circ}}$ for each Ge$_{1-x}$Sn$_x$ layer. Data have been mirrored to negative magnetic fields to better emphasize the Lorentzian character of the polarization decay. The inset shows the Π product ($g \cdot Ts$), obtained from FWHM of the Hanle curves, as a function of the Sn content. (b) Carrier lifetime $\tau$ as a function of Sn content, obtained by assuming a constant g-factor of 1.48 [19] and $T_s \approx \tau$. The error bar has been obtained by propagating the error of approximately 0.2 mT associated with measurement of the magnetic flux density. Inset: radiative lifetime $\tau_R$ estimated via the model theoretical calculation outlined in the Supporting Information

In the following, we introduce a magneto-optical investigation under CW excitation with the aim of elucidating the carrier dynamics in Ge$_{1-x}$Sn$_x$. Specifically, we performed Hanle measurements by sweeping the intensity of the magnetic field in a Voigt configuration, that is, along the direction orthogonal to the optically-defined spin orientation[19]. Figure 2(a) shows that for all the samples, $\rho_{circ}$ decays steadily with the strength of the external field. The data points have been normalized to the zero field value and henceforth mirrored to negative fields to better compare the expected Lorentzian line shape of the Hanle curves of the various Ge$_{1-x}$Sn$_x$ layers[20,33].

The FWHM of the Hanle curve is related via the Bohr-magneton $\mu_B$ and the reduced Planck constant $\hbar$ to the product P of the effective electron Landè g-factor $g$ and spin lifetime $T_s$ as[20,33]:

$$\Delta B = 2 \frac{\hbar}{g \mu_B T_s} = \frac{2\hbar}{\mu_B} \frac{1}{\Pi} \qquad [1]$$

Fig. 2(a) notably demonstrates the surprising impact of Sn incorporation on the Hanle effect. Very weak magnetic flux density of 20 mT, are indeed sufficient to suppress the PL polarization of the Ge$_{0.94}$Sn$_{0.06}$ layer [cyan curve in Fig 2(a)], yielding the narrowest Hanle linewidth, that is, the largest Π value of the samples investigated. The violet curve in Fig. 2(a) shows that the largest FWHM occurs for the epitaxial layer having the highest Sn molar fraction x = 0.1. The Ge$_{0.90}$Sn$_{0.10}$ film consequently possess the smallest Π. The non-monotonic dependence of the extracted values of Π on the Sn molar fraction is shown in the inset of Fig. 2(a). Such intrinsic characteristics with a pronounced maximum at x=0.06 can be ultimately ascribed to the alloy-induced changes of both the Landé g-factor and the spin lifetime.

It is worth noting, however, that all the coherent Ge$_{1-x}$Sn$_x$ epilayers studied in this work possess a dominant indirect bandgap character and a large built-in compressive strain. The optical transitions observed in our PL experiments thus allow us to gather access to the dynamics of spin-polarized electrons dwelling within the L-valleys even in Sn-richer alloys. According to the only available data, which suggest a $g$ value of 1.48 in Ge$_{0.95}$Sn$_{0.05}$[19], Sn incorporation in Ge is expected to smoothly decrease the g-factor of conduction band electrons from the pure Ge value of approximately 1.6[34–37]. This will result in a monotonic contribution to the $\Pi$ dependence, suggesting that the bell-shaped behavior shown in the inset of Figure 2(a) can be chiefly ascribed to the effect of alloying on the spin lifetime.

The latter is defined by the spin relaxation time $\tau_s$ and the carrier lifetime $\tau$ as $\frac{1}{T_s} = \frac{1}{\tau} + \frac{1}{\tau_s}$. Nonetheless, the sizeable $\rho_{circ}$ observed for all the samples under non-resonant excitation and in absence of the external magnetic field implies that $\tau_s > \tau$, and hence that $T_s \approx \tau$ [20]. This assumption was verified experimentally in a Ge$_{0.95}$Sn$_{0.05}$ sample exhibiting a spin relaxation time $\tau_s \sim 10\ ns$ and a carrier lifetime $\tau \sim 2\ ns$ [19]. The fit of the Lorentzian Hanle curves of Figure 2(a) therefore provides us with an estimate of the photogenerated carrier lifetime without the need to rely on time-resolved measurements.

Figure 2(b) shows the Sn-dependent $\tau$ values derived from $\Pi$ by assuming a constant g-factor $g = 1.48$ based on our recent experimental measurements[19]. The effective carrier lifetime is found to be in the hundreds of ps regime and approaches several of ns in the Ge$_{0.94}$Sn$_{0.06}$ layer. Remarkably, this result compares well with the lifetime modelled by using temperature dependent PL characteristics by Wirths et al.[38] and is fully consistent with the value of $\sim 2\ ns$, which was independently obtained by recent time-resolved PL investigations[19]. This further corroborates the effectiveness of our approach and supports the conclusion that our Hanle investigation can reliably yield information about the carrier kinetics in group IV materials.

We note that the measured carrier lifetime $\tau$ corresponds to an effective lifetime for photogenerated carriers, and encapsulates contributions from radiative ($\tau_R$) and non-radiative ($\tau_{NR}$) recombination processes: $\frac{1}{\tau} = \frac{1}{\tau_R} + \frac{1}{\tau_{NR}}$. Recent analysis[27,28] have shown that the evolution of the Ge$_{1-x}$Sn$_x$ bandgap is driven by Sn-induced hybridisation between the Ge Γ- and L-point conduction band edge states. A direct bandgap then emerges continuously with increasing Sn composition, as the alloy conduction band edge acquires increasing direct (Γ) character. On this basis, it is expected that the radiative lifetime $\tau_R$ decreases monotonically with increasing Sn composition, as the alloy bandgap acquires increasing direct character. To understand the measured trends in $\tau$ we perform a simple model estimate of the Sn composition dependent $\tau_R$, based on which we infer the role of non-radiative recombination. Our theoretical model – details of which are provided as Supporting Information – describes Sn-induced Γ-L mixing using a model Hamiltonian parametrised via atomistic alloy supercell electronic structure calculations. The lifetime $\tau_R$ associated with direct radiative recombination is then computed as

$$\tau_R(x) = \frac{E_g(0)\tau_R(0)}{E_g(x)f_\Gamma(x)} \qquad [2]$$

where $E_g(0)$ and $E_g(x)$ are the direct bandgaps of Ge and Ge$_{1-x}$Sn$_x$, respectively, $f_\Gamma(x)$ is the Ge Γ character of the Ge$_{1-x}$Sn$_x$ conduction band edge, and $\tau_R(0)$ is the radiative lifetime associated with the direct bandgap in Ge. Using our theoretical model, we compute $\tau_R(0) = 7.72$ ns, in good agreement with the value calculated from first principles by Rödl et al.[39] Our estimate of $\tau_R$, shown in the inset of Fig. 2(b), decreases with increasing Sn content, reaching a minimum value of 19.1 ns for x = 0.05. While this estimate neglects the impact of short-range substitutional alloy disorder on the band structure[28] and considers only zone-centre transitions[39], we note that these effects are expected to produce only minor quantitative changes to the computed $\tau_R$. We note that this theoretical estimate does not consider phonon-assisted recombination. The simultaneous requirement for electron-hole recombination and phonon emission or absorption in such processes reduces the associated recombination rate, thereby increasing the associated lifetime. As such, our calculated $\tau_R$ constitutes a lower bound on the radiative lifetime. On the basis that the calculated radiative lifetime is approximately one order of magnitude larger than the measured effective carrier lifetime, we conclude that carrier recombination in these samples is dominated by non-radiative recombination processes which occur sufficiently quickly to dominate the carrier kinetics.

Since the photogenerated excess carrier density is expected to be too low to give rise to significant Auger recombination[40], the likely scenario that explains the non-monotonic variation of $\tau$ with Sn content in Fig. 2(b) is related to the structural quality of the Ge$_{1-x}$Sn$_x$ layers and to the challenge of minimizing crystal defects during the CVD of the epitaxial solid thin film. We thus recall that strong out-of-equilibrium conditions have to be applied to incorporate Sn into the Ge lattice, i.e. low temperature (~200-300°C) and high growth rates to limit the surface diffusion of the adatoms. High Sn concentrations and strain values further decrease the Sn solubility[41,42]. This enhances the tendency of Sn segregation and can result in appearance of point defects and possible clustering of Sn. On the other hand, in the very diluted Sn regime the deposition temperature needs to be increased to approach the optimal growth temperature of elemental Ge. This eventually contributes to decease the Sn incorporation, in spite of the Ge-rich environment and low strain level of the target epitaxial layer. The competition of all these effects maximizes the structural quality of our coherent RP-CVD Ge$_{1-x}$Sn$_x$ films at an intermediate Sn composition x = 0.06, whose resulting superior optical properties manifest themselves through the lengthening of the effective carrier lifetime and give rise to the bell-shape behavior shown in Fig. 2(b).

Having addressed the carrier dynamics in coherently strained layers, in the following we extend the magneto-optical analysis to study partially relaxed materials. In so doing, we can obtain insight into the non-radiative recombination pathways introduced by the nucleation of extended defects due to plastic strain relaxation. Here we consider a set of samples having the same Sn molar fraction, namely $x = 0.08$. The thickness of the Ge$_{0.92}$Sn$_{0.08}$ epilayer was systematically varied across the critical thickness, i.e. between 40 and 80 nm, to precisely tailor strain relaxation R and to suitably inject a well-controlled dislocation density. Table 2 summarizes the experimental data obtained by x-ray diffraction (XRD) and transmission electron microscopy (TEM) measurements, showing that R ranges from 0 to 6%, while the linear dislocation density δ is between 0 and 2 μm$^{-1}$.

Table 2 Ge$_{0.92}$Sn$_{0.08}$ with different layer thickness h. The relaxation (R) has been experimentally determined by XRD data, and the linear dislocation density δ has been derived from the analysis of various TEM images. The strain relaxation of the 50 nm-thick sample is possibly below the experimental sensitivity, thus the linear dislocation density of this sample was estimated according to the procedure described in the Supporting Information.

| h (nm) | R (%) | δ (μm$^{-1}$) |
|---|---|---|
| 40 | 0 | 0 |
| 50 | 0 | 0.2 |
| 60 | 0.6 | 0.59 |
| 80 | 6 | 2.28 |

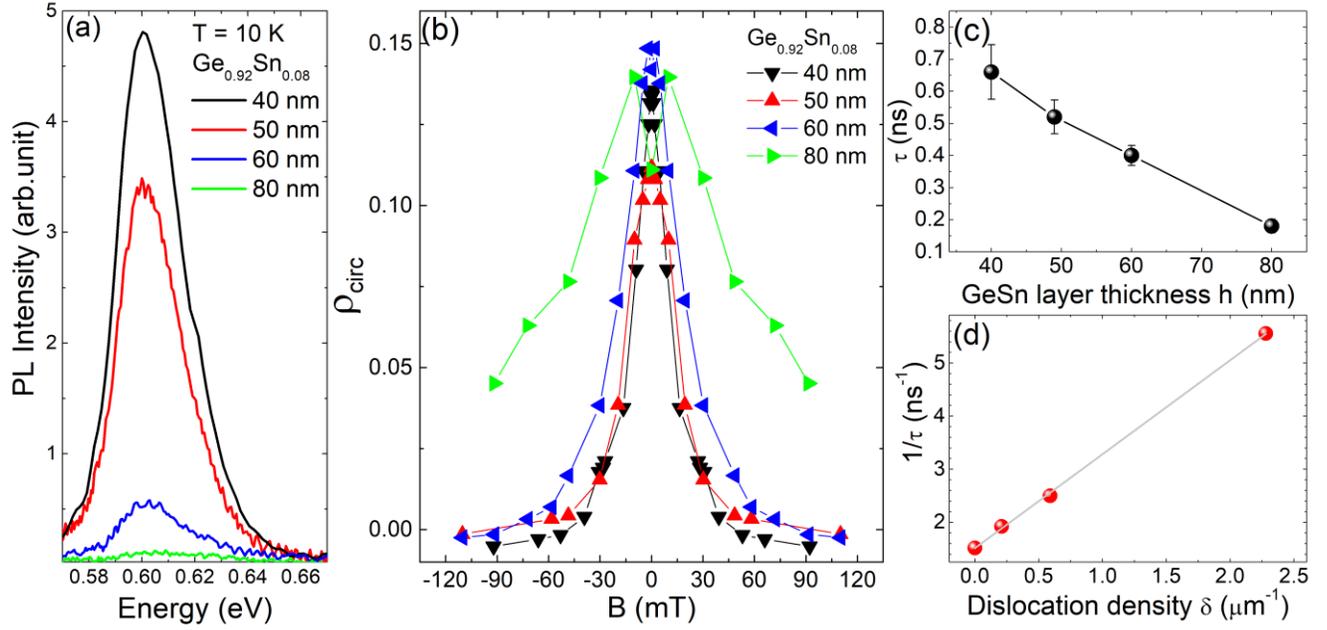

**Figure 3** (a) Low temperature PL spectra and (b) Hanle curves of the Ge$_{0.92}$Sn$_{0.08}$ layers with different thicknesses. Hanle data have been mirrored at negative magnetic fields. (c) Carrier lifetime $\tau$ as a function of layer thickness obtained from the Hanle decay curves. (d) Inverse of carrier lifetime as a function of the linear dislocation density. The linear fit shows an increase with a slope of $1.77 \pm 0.03 \cdot 10^5$ cm/s.

Figure 3(a) shows the low temperature PL spectra of the Ge$_{0.92}$Sn$_{0.08}$ samples. We find that the PL intensity scales inversely with the epilayer thickness, so that the thickest Ge$_{0.92}$Sn$_{0.08}$ film demonstrates the weakest PL signal[43]. The suppression of the emission mimics the strain relaxation process and reveals the introduction of non-radiative recombination pathways associated with the nucleation of dislocations at the interface between the Ge$_{1-x}$Sn$_x$ layer and underlying Ge buffer. Relaxation process occurs via formation of dislocations at the Ge$_{1-x}$Sn$_x$/Ge interface with subsequent roughening of the Ge$_{1-x}$Sn$_x$ epilayer's surface.

The Hanle curves reported in Fig. 3(b) clarify that by increasing the epilayer thickness h and, thus, the dislocation density δ, the Hanle curve broadens, implying a reduction of the effective carrier lifetime $\tau$. The $\tau$ values that were derived from the magneto-optical data are then summarized in Fig. 3(c). The data points demonstrate the compelling link between the effective carrier lifetime and the layer thickness. Similar effects on the carrier lifetime have been observed in strained III-V heterostructure[44].

We now elaborate further on this finding to provide a refined description of the kinetics that takes place in presence of an increasing number of extended defects that can act as carrier sinks. At first, we notice that recombination at the free surface provides a competing non-radiative channel. Yet, this has the same weight in the carrier dynamics of all the studied samples. The native oxide of the un-passivated surface cannot be modified by the mild strain relaxation of our epitaxial films. In addition, the laser penetration depth and the carrier diffusion guarantee a homogeneous distribution of the photogenerated carriers through the whole thickness of all the Ge$_{0.92}$Sn$_{0.08}$ layers[45,46]. As a result, changes in the non-radiative lifetime of the epitaxial films can be ascribed to the emergence of recombination at the defective Ge$_{1-x}$Sn$_x$/Ge heterointerface. Specifically, we plot in Fig. 3(d) the measured carrier recombination rate $1/\tau$ as a function of linear dislocation density as derived from the TEM analysis summarized in Tab. 2 (see the Supporting Information for more details). Fig. 3(d) notably demonstrates a linear behavior. Such a result agrees well with the findings obtained in Si and Ge by Kurtz et al[47,48]. The slope of the line that fits the data points shown in Fig. 3(d) provides us with a relevant experimental estimation of $1.77 \pm 0.03 \cdot 10^5 \ cm/s$ for the recombination velocity at dislocations in $Ge_{1-x}Sn_x$. Remarkably, such value is approximately two orders of magnitude larger than previous reports for dislocated Si-rich SiGe/Si heterointerfaces[49,50] and between two and ten times larger than in junctions between Si and elemental Ge[51,52]. This should be carefully considered in optimizing ultrathin $Ge_{1-x}Sn_x$ heterostructure for future optoelectronic applications. Also, for fully strained $Ge_{1-x}Sn_x$ epilayers, our characterization methodology based on magneto-optics demonstrates to be an accurate and reliable approach to improve and optimize the structural and optical quality of the material.

In conclusion, we have introduced an effective approach for the determination of carrier lifetime that has been largely overlooked in the context of group IV materials. By taking advantage of well-developed steady-state magneto-optics, based on the Hanle effect, we unveiled a sub-ns lifetime for the photogenerated carriers in coherently strained $Ge_{1-x}Sn_x$ epilayers. The investigation has been extended up to a Sn content of 10%. Our findings suggest that the carrier kinetics is dominated by the presence of non-radiative defect-related recombination associated with imperfection of the $Ge_{1-x}Sn_x$ epitaxial material, arising due to the challenging epitaxy of $Ge_{1-x}Sn_x$ by CVD at reduced pressure. By increasing the thickness of $Ge_{1-x}Sn_x$ layers, we have been able to promote and investigate the consequences of well-controlled plastic strain relaxation. We demonstrated the emergence of the parasitic non-radiative recombination associated with the concomitant nucleation of dislocations. Owing to the fast recombination velocity at the defective $Ge_{1-x}Sn_x$ /Ge interface, the largest the dislocation density corresponds to the shortest measured effective carrier lifetime. Our findings inform fundamental investigations of the carrier dynamics in novel group IV alloys, and provide key information for the future design of $Ge_{1-x}Sn_x$-based devices, driving novel silicon photonic concepts and optoelectronic device design.

ACKNOWLEDGMENTS

The authors thank L. Franzini for technical assistance. This work was supported by Science Foundation Ireland (SFI; project no. 15/IA/3082). CAB acknowledges the support of the National University of Ireland Post-Doctoral Fellowship in the Sciences.